\documentclass{llncs}

\usepackage{amsmath, amsfonts, amssymb, graphicx, float}
\usepackage{amsbsy}
\usepackage{mathrsfs}
\usepackage{bm}
\usepackage{array,booktabs}
\usepackage{verbatim}
\usepackage{multirow}
\usepackage{hyperref}
\usepackage{subfigure}

\newcommand{\avg}[1]{\left\langle #1 \right\rangle}

\begin{document}

\title{Structural Properties of Ego Networks
\thanks{This  work was supported in part by AFOSR (contract FA9550-10-1-0569), by NSF (grant CIF-1217605)  and by DARPA (contract W911NF-12-1-0034).}}
\author{Sidharth Gupta\inst{1}, Xiaoran Yan\inst{2}, Kristina Lerman\inst{2}}
\institute{Indian Institute of Technology, Kanpur, India
\and Information Sciences Institute, University of Southern California, USA
}
\maketitle

\begin{abstract}
The structure of real-world social networks in large part determines the evolution of social phenomena, including opinion formation, diffusion of information and influence, and the spread of disease. Globally, network structure is characterized by features such as degree distribution, degree assortativity, and clustering coefficient. However, information about global structure is usually not available to each vertex. Instead, each vertex's knowledge is generally limited to the locally observable portion of the network consisting of the subgraph over its immediate neighbors. Such subgraphs, known as \emph{ego networks}, have properties that can differ substantially from those of the global network. In this paper, we study the structural properties of ego networks and show how they relate to the global properties of networks from which they are derived. Through empirical comparisons and mathematical derivations, we show that structural features, similar to static attributes, suffer from paradoxes. We quantify
the differences between global information about network structure and local estimates. This knowledge allows us to better identify and correct the biases arising from incomplete local information.
\end{abstract}

\section{Introduction}

As powerful representations for complex systems, networks model entities and their interactions as vertices and edges.
Over the years, different attributes characterizing real world networks have been proposed and investigated. These include features like the degree distribution, degree assortativity and clustering coefficient that describe network structure at the global level. Many efficient models and algorithms have been developed for their generation and inferences.~\cite{albert2002statistical,newman_structure_2003}.

Unfortunately, efficient algorithms usually rely on global knowledge of the network, which is typically not available to each vertex of the network. This is especially the case for real world social networks like the one in Milgram's ``small world'' experiment~\cite{milgram1967small}. In social networks where vertices correspond to people, without digital bookkeeping, individuals only have access to \emph{local information} about their immediate neighbors. Even in online networks such as Facebook, information access is restricted by privacy settings.

While small world structures can generally explain efficient decentralized navigation~\cite{kleinberg_complex_2006}, other connections between global and local network measures are less understood. The ``friendship paradox,'' for example, states that on average, your friends have more friends than you do~\cite{feld_why_1991}, which can be generalized to many other attributes ~\cite{hodas_friendship_2013,kooti_network_2014}. These systematic biases have been widely observed in social studies, for attributes ranging from wealth~\cite{amuedo2007social} to epidemic risk~\cite{christakis2010social}, and has been largely attributed to distribution bias in the sampling process. These paradoxes can at a local level distort our perceptions of the ground truth, resulting in inefficient policies and social consequences.

Local information about network structure from the perspective of a vertex is captured by its ego network, which is the induced subgraph over that vertex's immediate neighbors. Ego networks are considered to be the basic structure that dominates the central vertex's perspectives and activities~\cite{newman_ego-centered_2003,leskovec_learning_2012,ugander_subgraph_2013,backstrom_romantic_2014}. In this work, we study the structural properties of ego networks and relate them to those of the global network. We hope to quantify structural biases arising from local, incomplete information, and recover accurate global estimates.

We will first review related work on the structural properties of global networks. In section 3.1, we will establish the mathematical mappings for degree distributions between the global and ego networks. In section 3.2, we will investigate degree assortativity and clustering coefficient at ego network level, by combining our knowledge of global structures and their mathematical relations.
%we hope to identify the mathematical and social origins of network paradoxes. With such knowledge, we will be able to

%In this work, we explore the origin of these paradoxes by comparing the structural features of global networks and their decomposed ego networks. The ego network of a vertex is the subgraph over its immediate neighbors, and is considered to be the basic local structure that dominates the central vertex's perspectives and activities~\cite{leskovec_learning_2012,backstrom_romantic_2014}. By studying the mathematical connections between the structural features of global and ego networks, and building generative models that can reproduce statistics at both levels, we hope to identify mathematical and social origins of these paradoxes. With such knowledge, we will be able to mitigate or even correct the biases from local and limited information, and recover better estimates of the global ground truth.

\section{Background and related work}
With traditional independent data, the global statistics of a population remain unbiased estimates for subsets. In networks, however, the complex dependencies can skew localized statistics, leading to inhomogeneity at different scales and positions. Numerous efforts have been made to develop generative models which can reproduce realistic structure with simple local algorithms~\cite{barabasi_emergence_1999,newman_mixing_2003,serrano_tuning_2005,clauset_power-law_2007}. Unfortunately, structural features are so intertwined that preserving one often biases another. The same difficulty is also observed in graph sampling, where different sampling techniques can lead to different biases~\cite{leskovec_sampling_2006,clauset_accuracy_2004}.

However, real world networks do exhibit certain patterns. We focus on comparing the collective perceptions of individuals with the global ground truth, similar to previous work for static features, where the bias in the sampling process leads to paradoxes~\cite{feld_why_1991,kooti_network_2014}. Structural features, unlike their static counterparts, change values over specific subgraphs, and thus have an additional complication.

In this section, we review and organize the relevant work on structural features of networks, including degree distribution, degree assortativity and clustering coefficient. They describe network structure at the global level. However, they are by definition aggregations of local measures, and they are closely related to each other. We will focus on undirected graph $G = (V,E)$ with vertex set $V$, edge set $E$, and size $N= |V|$.

Degree distribution is one of the best studied aspects of networks. Many real world networks display ``scale-free"~\cite{barabasi_emergence_1999,clauset_power-law_2007}, or power law degree distributions:
\begin{equation}
d_u \sim P(k) = \frac{\gamma-1}{k_{min}}(\frac{k}{k_{min}})^{-\gamma}\;,
\end{equation}
where $k=d_u$, $k_{min} = d_{min}$ and the range of the distribution is $[d_{min},\infty]$. The exponent $\gamma$ is usually in the range $[1,3]$. 

%The degree distribution $P(d_u)$ is a powerful statistical tool capturing population proportions of vertices based on their local connectivities. An understanding of this distribution is essential for the study of higher order structural features, and multiple models have been proposed to generate graphs with given degree distributions.

While specifying individual degrees is a step forward from simple random graph models, real world networks also exhibit higher order correlations. Degree assortativity, for example, captures the pair-wise correlation between the degrees of neighboring vertices~\cite{newman_mixing_2003}. In fact, mathematical constraints alone can predict that scale-free networks with $\gamma<3$ cannot be completely uncorrelated, leading to the phenomenon of ``structural cut-off"~\cite{boguna_cut-offs_2004} --- smaller the value of $\gamma$, lower the maximum possible positive correlation between the degrees of neighbors in the network. Many real world networks are thus more disassortative than we would normally expect. 

Degree correlations are fully specified by the joint distribution:$e(k, k') = \frac{E(k,k')}{\avg{k}N}$, where $E(k,k')$ is the number of edges between vertices of degree $k$ and $k'$.
\begin{comment}
Given the degree distribution $P(k)$, we can calculate the average degree of nearest neighbors (ANND) for vertices of a certain degree~\cite{pastor2001dynamical}:
\begin{equation}
 k_{annd} (k) = \sum_{k'} k'\frac{e(k, k')}{q(k)} = \sum_{k'} \frac{k'E(k,k')}{kN(k)}\;,
\end{equation}
where $q(k)$ is the probability of a degree $k$ vertex is sampled by following edges:
\begin{equation}
q(k) = \frac{kP(k)}{\avg{k}} =\frac{kN(k)}{\avg{k}N}\;,
\end{equation}
and $N(k)$ is the number of vertices of degree k.

When $k_{annd}(k)$ is an increasing function of k, we say the global network is assortative.
\end{comment}
A scalar aggregation of local assortativity in the way of Pearson's correlation gives us the global assortativity,
\begin{equation}
\label{eq:assortativity}
r_{glo} = \frac{1}{\sigma^2_q} \sum_{k,k'}kk'[e(k,k')-q(k)q(k')]\;,
\end{equation}
where $q(k) = \frac{kP(k)}{\avg{k}} = \frac{kN(k)}{\avg{k}N}$ is the probability of sampling a vertex of degree $k$ by following a randomly chosen edge and $\sigma^2_q$ is the variance of $q(k)$.

The complexity of real world networks does not stop at pair-wise correlations. Clustering coefficient goes one step further, capturing correlations among triplets of vertices~\cite{watts1998WS}. The local version is defined as the probability that a third edge between two neighbors of the same vertex $v$ would complete a triangle, with $C_v= \frac{2T_v}{d_v(d_v-1)}$, where $T_v$ is the number triangles containing the vertex $v$. We can aggregate $C_v$ over the set of vertices of a given degree $d_v = k$, and get the degree dependent clustering coefficient~\cite{vazquez2002large},
\begin{equation}
C(d_v)= C(k) = \frac{1}{N(k)} \sum_{v, d_v=k} C_v = \frac{1}{N(k) k(k-1)}\sum_{v, d_v=k} 2T_v\;,
\end{equation}
where $N(k)$ is the number of vertices of degree k. In real world networks, it has been observed that $C(k)$ is also a power law function of degree,
$C_{d_u} = C_0 d_u^{-\alpha}\;,$
where $\alpha$ typically ranges from $[0,1]$, with networks having strong hierarchical structures corresponding to $\alpha=1$~\cite{ravasz_hierarchical_2003}. $C_0$ is a constant depending on global clustering coefficient $C_{glo}$. Given the degree distribution $P(k)$, we can recover
$C_{glo} = \sum_{k=2}^{k_{max}}P(k) C(k)\;,$
where we only consider vertices with $k>1$.

Being a third order correlation measure, clustering coefficient displays dependencies on both degree distribution and degree correlations or assortativity~\cite{boguna_class_2003}. The interplay between degree correlations and clustering is further complicated by the fact that each edge can form multiple triangles. It has been shown that negative degree correlations can limit the maximum value of $C_{glo}$, as triangles are less likely to appear with disassortative connections \cite{serrano_tuning_2005}.

\begin{comment}
If we define the number of triangles sharing the edge $(u,v)$ as $m_{uv}$ (also called the edge multiplicity or embeddedness in earlier works), we can show that
\begin{equation}
 \sum_{d_u=k}\sum_{d_v=k'} m_{uv} = m(k,k') E(k,k') = m(k,k') e(k,k')\avg{k}N\;,
\end{equation}
where $m(k,k')$ is the average multiplicity of edges connecting degree $k$ and degree $k'$ vertices.

Since we know the following identity
\begin{equation}
\label{eq:multiplicity-clustering}
 \sum_v m_{uv} = 2T_u = C_ud_u(d_u-1)\;,
\end{equation}
we can sum over vertices of degree $k$,
\begin{align*}
 &\sum_{d_u=k} \sum_{k'}\sum_{d_v=k'} m_{u,v} = \sum_{d_u=k} C_ud_u(d_u-1)\\
  &= \sum_{k'} m(k,k') e(k,k')\avg{k}N = N(k)C(k)k(k-1)\;.
\end{align*}

By summing over the other index $k$, we recover the average edge multiplicity of the network $m_{glo}$~\cite{serrano_tuning_2005}:
\begin{equation}
\label{eq:avg-multiplicity}
 \sum_{k,k'} m(k,k') e(k,k') = \sum_k \frac{P(k)}{\avg{k}}C(k)k(k-1) = m_{glo}\;.
\end{equation}
The above equation and the concept of edge multiplicity will play an important role in our analysis of ego network structures.
\end{comment}

\section{Structural features of ego networks}

\begin{figure}[tbh]
\begin{center}
\begin{tabular}{cc}
%\subfigure[Karate club network] {
%\label{fig:karate-global}
\includegraphics[width=0.4\textwidth]{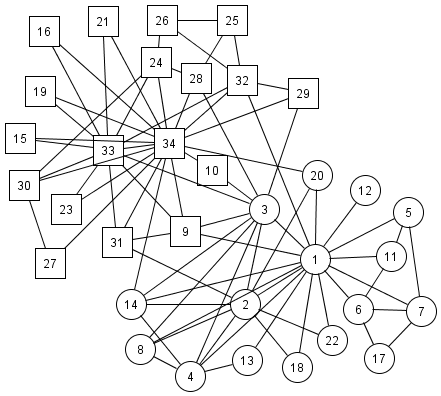}%}
&
%\subfigure[Ego network of vertex 33] {
%\label{fig:karate-ego}
\includegraphics[width=0.4\textwidth]{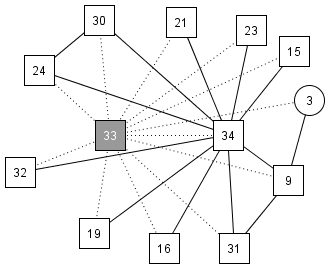}
%}
\\ (a) & (b)
\end{tabular}
\caption{A benchmark social network representing friendships among members of a karate club~\cite{zachary} (a) at the global level and (b) for the ego network of vertex 33.}
\label{fig:karate}
\end{center}
\end{figure}

An ego network is defined as the subgraph induced over vertices directly connected to a specific vertex, called an ego, but excluding the ego itself~\cite{ugander_subgraph_2013,backstrom_romantic_2014}. A toy example is given in Figure~\ref{fig:karate}. Keep in mind that the removal of the ego can disconnect an ego network.

By definition, ego network structure is closely connected to the local structure around the ego in the global network. These relationships are important to our understanding of perceptions based on local and limited information. In this section, we investigate how degree distribution, degree assortativity and clustering coefficient of ego networks depends on those of the global network. Although a full generative model that reproduces all structural features at both global and ego networks levels is difficult to build, we can however leverage our knowledge of global structures. Combining that with mathematical mappings between the two levels, we can better understand and even predict structures in ego networks.

We will approach the problem with theoretical derivations. To keep the mathematics tractable and intuitive, we will make some simplifying assumptions and educated guesses during the process. Therefore, it is very important to support our claims with empirical evidence. Our studies span a diverse range of network datasets where vertices correspond to people, such as social, coauthorship, communication and hybrid (serving social and informational purposes) networks~\cite{snapnets}, as detailed in Table~\hyperref[tab:dataset_description]{\ref*{tab:dataset_description}}.

\begin{table}
\caption{Description of the network datasets used for empirical studies}
\begin{center}
	\begin{tabular}{|l|l|l|l|l|l|l|}
	\hline
	Dataset 			& Type 		&$|V|$ 		& $|E|$ 		& $90\%$ Eff 	& $r_{glo}$ & $C_{glo}$\\
					&		&		&		 	& Diameter 	&	&\\	
	\hline
	Facebook		& Social		& 4,039		& 88,234 		& 4.6 		& 0.064	& 0.61\\
	\hline	
	Orkut			& Social 		& 3,072,441	& 117,185,083 		& 4.8		& 0.016	& 0.17\\
	\hline	
	General Relativity	& Coauthorship 		& 5,242		& 14,496 		& 7.9		& 0.66	& 0.53\\
	\hline
	High Energy Physics	& Coauthorship		& 12,008	& 118,521 		& 5.7		& 0.63	& 0.61\\
	\hline
	Enron email		& Communication		& 36,692	& 183,831 		& 4.7		& -0.11	& 0.50\\
	\hline
	LiveJournal		& Hybrid		& 3,997,962	& 34,681,189 		& 6.5		& 0.045	& 0.28\\
	\hline
	\end{tabular}
\end{center}
\label{tab:dataset_description}
\end{table}

In the following subsection, we will treat all ego networks as a giant disconnected graph. Here a vertex $u$ will appear $d_u$ times, and we index the features of each instance using a superscript. For example, the degree of vertex $u$ in the ego network of $v$ is denoted as $d^v_u$. Features without upper indices are global measures.

\subsection{Degree Distribution}
We start off by investigating the degree distribution in ego networks. The first simple connection to observe is that the size of the ego network of vertex $v$ is simply its global degree $d_v$. The edge density of the ego network of vertex $v$ is the local clustering coefficient $C_v$ in the global network. The degree of vertex $u$ in the ego network of $v$, or $d^v_u$, corresponds to the number of triangles containing the edge $(u,v)$, which is symmetric for undirected graphs
\begin{equation}
d_u^v = d_v^u = m_{uv}\;,
\end{equation}
where $m_{uv}$ is the number of triangles sharing the edge $(u,v)$. By summing over egos, we get the total degree of vertex $u$ across all ego networks that it appears in, which is equal to the total degree of all vertices in the ego network of $u$:
\begin{equation}
 \sum_v d_u^v = \sum_v d_v^u = 2T_u = C_ud_u(d_u-1)\;.
\end{equation}
The above equality gives us the average degree of vertex $u$ across ego networks:
\begin{equation}
\avg{d^v_u} = \frac{\sum_v d^v_u}{d_u} = C_u (d_u-1)\;,
\end{equation}
which by symmetry is also the average degree of the ego network of vertex $u$. However, if we treat each instance of vertex $u$ in all ego networks as independent variables, the average becomes:
\begin{equation}
 \avg{d^v_u}_{ego} = \frac{d_u\avg{ d^v_u }}{\avg{d_u}} = \frac{1}{\avg{d_u}} C_ud_u(d_u-1)\;.
\end{equation}
This over-representation of high degree vertices is the result of edge sampling. If we assume that both the global degree distribution and the degree dependent clustering follow power laws, as defined in the previous section, we get
$$
d_u \sim P(x) = \frac{\gamma-1}{x_{min}}(\frac{x}{x_{min}})^{-\gamma}\;, \quad\quad C_{d_u} = C_0 d_u^{-\alpha}\;,
$$

\begin{comment}
$\avg{d^v_u}_{ego}$ also has an interesting connection with the average edge multiplicity of the global network (Eq. \ref{eq:avg-multiplicity}):
\begin{align}
\label{eq:meanAsMultiplicity}
 \sum_u \avg{d^v_u}_{ego} =& \sum_k\sum_{d_u=k} \frac{1}{\avg{k}} C_ud_u(d_u-1)\nonumber\\
		      =& \sum_k  \frac{N(k)}{\avg{k}}  C(k)k(k-1) = N m_{glo}\;.
\end{align}
Or in another word, the arithmetic mean of $\avg{d^v_u}_{ego}$ is the average edge multiplicity of the network.
\end{comment}

By a change of variables $\avg{ d^v_u }_{ego} \approx \frac{1}{\avg{d_u}} C_u d_u^2 = \frac{C_0}{\avg{d_u}} d_u^{(2-\alpha)} = Z d_u^{(2-\alpha)}$, we have
\begin{align}
\label{eq:changeVR}
\avg{d^v_u}_{ego} \sim P(y) &=  \left| \frac{1}{2-\alpha} (\frac{y}{Z}) ^{\frac{\alpha-1}{2-\alpha}} \frac{1}{Z}\right|
				    (\frac{y}{Z})^{\frac{-\gamma}{2-\alpha}} x_{min}^{\gamma} \frac{\gamma-1}{x_{min}}\nonumber\\
	      &= \frac{x_{min}^{(\gamma-1)}(\gamma-1)}{(2-\alpha)Z} (\frac{y}{Z}) ^{\frac{\alpha-\gamma-1}{2-\alpha}}\;.
\end{align}

Since most real world networks have $1\leq\gamma\leq 3$ and $0\leq\alpha\leq 1$, the above exponent can be written as
$$
\frac{\alpha-\gamma-1}{2-\alpha} = -\gamma + \frac{(\gamma-1)(1-\alpha)}{2-\alpha} \geq -\gamma\;,
$$
which means the $\avg{ d^v_u }_{ego}$ actually follows a power law with a heavier tail than the original degree distribution $P(k)$. In the extreme when $\alpha=1$, as in many cases for networks with strong hierarchical structures~\cite{ravasz_hierarchical_2003}, the mean degree of vertex $u$ across ego networks and the mean degree of ego networks both become constants (uniform distribution).

The full distribution of $d^v_u$ generally requires the complete knowledge of higher correlations. However, we do know that by definition $\avg{d_u^v}=E[d^v_u] = E[\avg{ d^v_u }_{ego}]$. Assuming it also follows a power law distribution, we have
\begin{align*}
d^v_u \sim P(z) =& \frac{\eta-1}{z_{min}}(\frac{z}{z_{min}})^{-\eta}\;,\\
E[d^v_u]=z_{min}(1+\frac{1}{\eta-2}) =& y_{min}(1+ \frac{1}{\frac{-\alpha+\gamma+1}{2-\alpha}-2}) = E[\avg{d^v_u}_{ego}]\;.
\end{align*}
Since smallest instances of $d^v_u$ is smaller than its average $\avg{d^v_u }_{ego}$, we have $z_{min} < y_{min}$ and thus $\eta< \frac{-\alpha+\gamma+1}{2-\alpha}$, which means that the full distribution of $d^v_u$ has a even heavier tail. Considering that $P(y)$ will be the same as $P(z)$ if all the vertex instances have the same degree, and we will underestimate the variance otherwise, we do expect $P(z)$ to have a wider spread.

\begin{figure}
\centering
\includegraphics[width=0.49\linewidth]{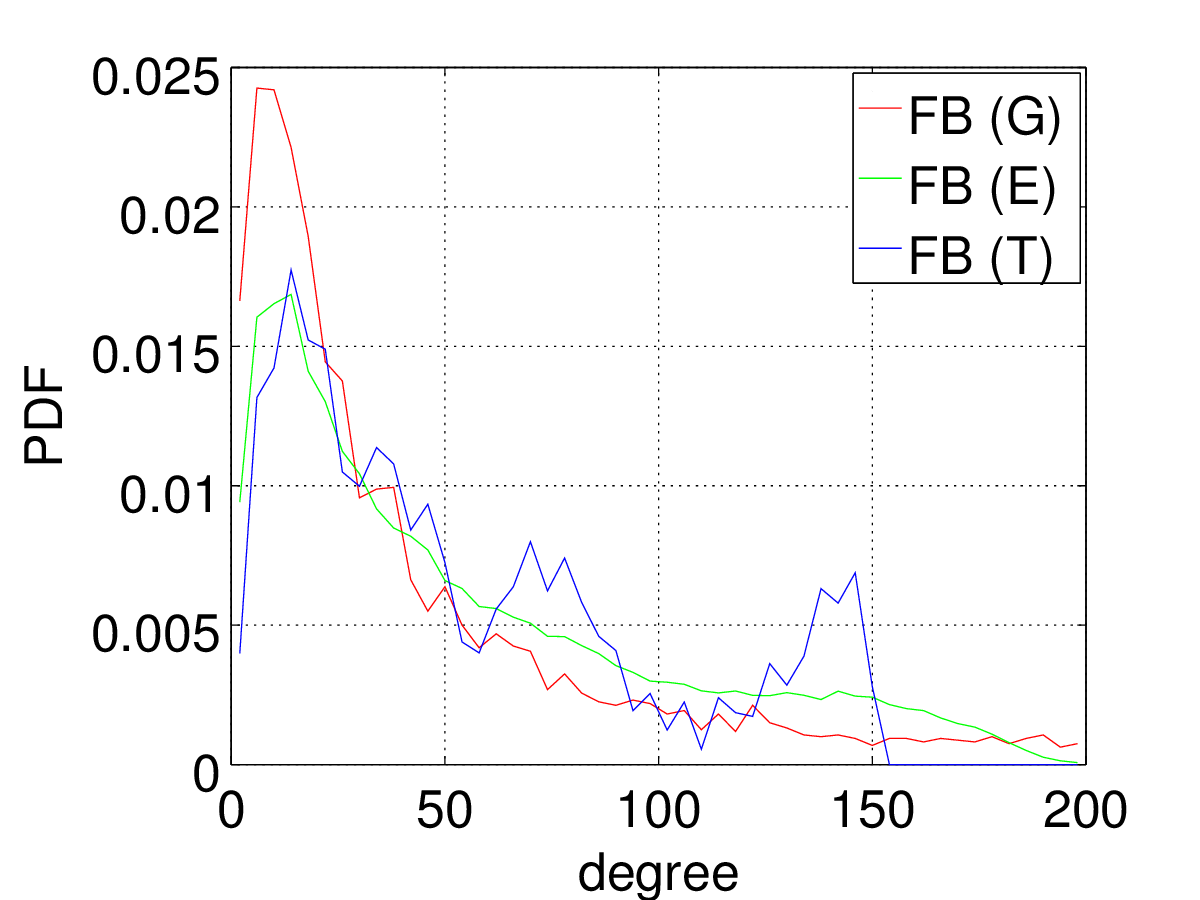}
\includegraphics[width=0.49\linewidth]{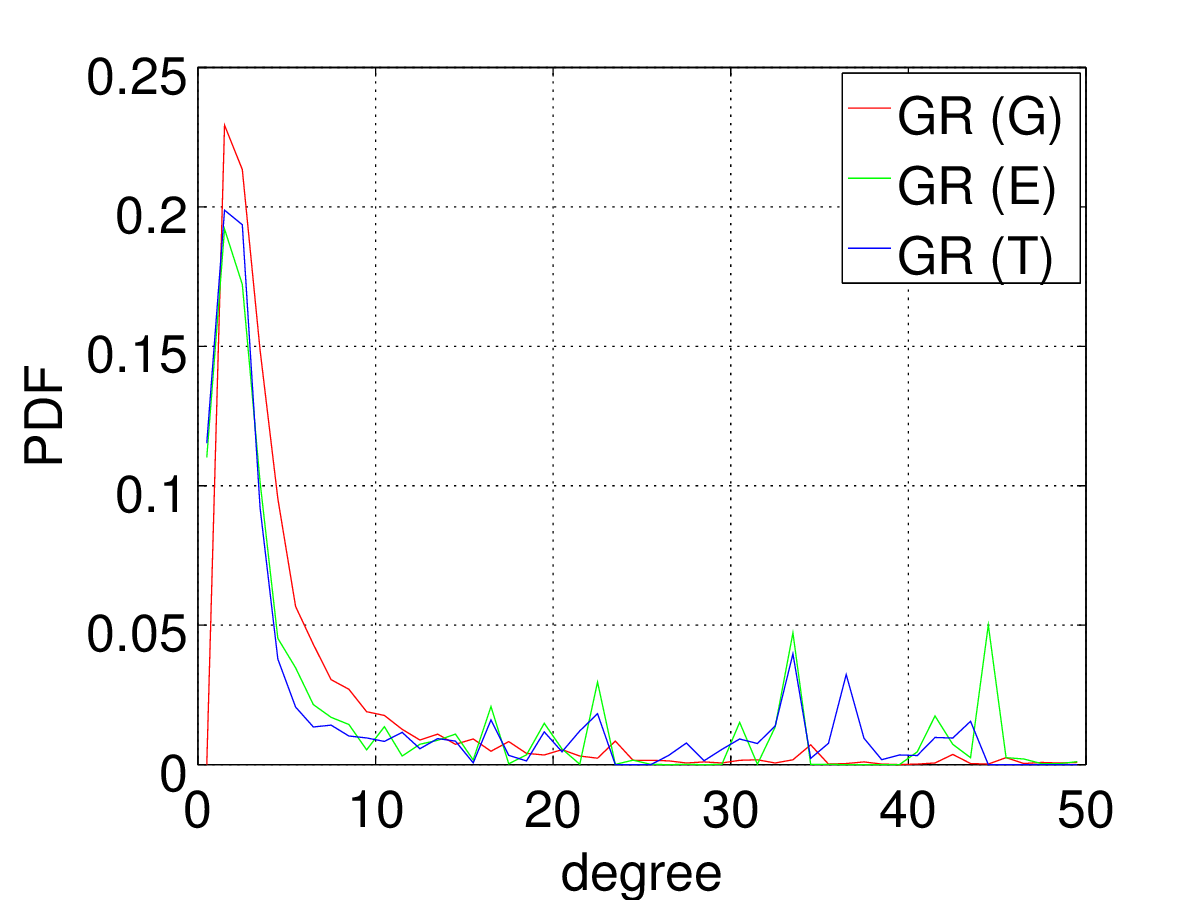}
\caption{Degree distributions of Facebook (left) and General Relativity (right). Red curves are for global degrees (G), green curves are for ego network degrees (E) and blue curves are for our theoretical approximation $\avg{d^v_u}_{ego}$ (T).}
\label{fig:degreeDist}
\end{figure}

This is consistent with our empirical observations (Figure.~\ref{fig:degreeDist}). In practice, the distribution of $\avg{d^v_u}_{ego}$ can be empirically constructed by putting $d_u$ copies of $C_u(d_u-1)$ together. Independent of the shape of $P(d_u)$, our intuitions that ego networks have heavier tails holds as long as $\alpha\leq 1$.

\begin{table}[H]
\caption{Properties of degree distributions at global and ego network levels}
\begin{center}
	\begin{tabular}{|l|l|l|l|l|l|l|l|l|}
	\hline
	\emph{Network} 			& $med(d_u)$ 		& $\avg{d_u}$ & $\avg{d_u}_{nn}$ & $\avg{d_u^v}$ & $\avg{d_u^v}_{nn}$ 	 &$P_{glo}$	& $frac_{u,v}(d_u^v = 0)$\\
	\hline
	Facebook				& 25			& 43.7		& 106.6 	& 54.6 		& 95.8 		& 1.1E-2	& 0.09\%\\
	\hline	
	Orkut					& 45 			& 76.3		& 390.3		& 16.1		& 72.3 		& 2.5E-5	& 13.64\%\\
	\hline	
	General Relativity			& 3 			& 5.5		& 16.9 		& 10.0		& 29.2 		& 1.1E-3	& 10.99\%\\
	\hline
	High Energy Physics			& 5			& 19.7 		& 129.9		& 85.0		& 187.0 	& 1.6E-3	& 2.16\%\\
	\hline
	Enron email				& 3			& 10.0 		& 140.1		& 11.9		& 34.5 		& 2.7E-4	& 7.65\%\\
	\hline
	LiveJournal				& 6			& 17.3		& 123.7 	& 15.4		& 149.1 	& 4.3E-6	& 16.75\%\\
	\hline
	\end{tabular}
\end{center}
\label{tab:degree_comparison}
\end{table}

Table~\hyperref[tab:degree_comparison]{\ref*{tab:degree_comparison}} summarizes empirical properties of degree distributions for all the data set. As predicted by our theory, $\avg{d_u^v}$ is usually greater than $\avg{d_u}$, except for two large networks Orkut and LiveJournal. Their low densities $P_{glo}$ lead to disconnected ego networks, illustrated by high fraction of degree zero ego network instances $frac_{u,v}(d_u^v = 0)$, which breaks our approximation in Eq.~\ref{eq:changeVR}.

In real social settings, the consistent bias of ego networks towards higher degrees can lead to wrong perceptions. For static features, over-representation of high degree hubs is identified as an important origin of ``friendship paradox" and its generalizations~\cite{feld_why_1991}. The heavy tail degree distributions make the matter worse if the arithmetic mean is used~\cite{kooti_network_2014}. Our analysis of degree distributions of ego networks show that both effects are still in play for structural features, leading to the surprising result $\avg{d_u^v}>\avg{d_u}$ even after the ego is taken out.

As a result, one should take extra caution when making claims about the global structure from local observations. Many popular connectivity and centrality measures for networks aggregates local structural features, they are thus potentially biased estimates of the global truth.  However, our derivation of $\avg{d^v_u}_{ego}$ also shows that with appropriate assumptions, we can approximate global truth by its mathematical connection to local measures. In this case, we suggest using $\avg{d^v_u} = C_u(d_u-1)$ to avoid over-representation, taking $C_u$ in to account when your information is limited to $u$'s neighborhood, and using medians instead of mean as suggested in~\cite{kooti_network_2014}.

\subsection{Degree assortativity and clustering coefficient}
In the global network, degree correlations are heavily constrained by the degree distribution. With our understanding of the ego network degree distribution, we are ready to study its implications. According to Eq.~\ref{eq:assortativity}, the assortativity of ego networks can be defined as
\begin{equation}
 r_{ego} = \frac{1}{V[d^v_u]}\left(\sum_{k,k' = min(d^v_u)}^{max(d^v_u)}kk'e_{ego}(k,k')- E^2[d^v_u] \right)\;,
\end{equation}
where we plugged in ego network level features. Assortativity is largely determined by the difference between the positive terms $kk'e_{ego}(k,k')$ and the negative terms $E^2[d^v_u]$. By the results of last subsection, we know that $E^2[d^v_u]$ has generally become bigger. For the former, if we again assume that all the instances of a degree $k$ vertex have the same degree $C(k)k$, we have
$$e_{ego}(kC(k),k'C(k')) = \frac{m(k,k')}{m_{glo}}e_{glo}(k,k')\;.$$
The change of the positive term depends on the details of $m(k,k')$, i.e. how edges are shared between triangles.
Our empirical observations, however, confirms that degree assortativity are smaller in ego networks (see Table~\hyperref[tab:clustering_comparison]{\ref*{tab:clustering_comparison}}). The reduction in degree assortativity across ego networks is consistent with the argument of structural cut-offs. Since we know that ego networks generally have fatter tails and thus smaller $\gamma$, they are naturally more disassortative.

Next we analyze clustering coefficients of ego networks. Based on our knowledge of global features, clustering coefficients have very complicated dependencies with degree distributions and assortativities. However, our empirical measure reveals a very simple pattern (see Table~\hyperref[tab:clustering_comparison]{\ref*{tab:clustering_comparison}}).

As compared to global networks, ego networks display only slightly higher clustering coefficients. If we consider the ego network of vertex $v$ a Erd\"{o}s\textendash R\'{e}nyi random graph, then the local clustering coefficient $C_v$ in the global network is the edge generating probability. Averaging it over all vertices we get $C_{glo}$. For the global network, this probability $P_{glo}$  is orders of magnitude smaller than $C_{glo}$. The insignificant difference between $C_{glo}$ and $C_{ego}$ indicates that ego networks are much closer to random graphs than global networks. This observation confirms what Ugander et al. reported in their study of subgraph frequencies~\cite{ugander_subgraph_2013}, where generative models with triangle closure is capable of reproducing higher order correlations observed in real world networks.

In fact, the probability of completing a triangle $(i,j,k)$ given the edges $(i,j)$ and $(j,k)$, in the ego network of $v$, is equivalent to the probability of completing the 4-clique $(i,j,k,v)$ in the global network, given the triangles $(i,j,v)$ and $(j,k,v)$. Assuming triangle completions at the global level are independent, with uniform probability $C_{glo}$, we can estimate $C_{ego}^{rand}$ for ego network clustering,
$$C_{ego}^{rand}= 1 - (1-C_v)(1-C_u) \approx 2 C_{glo} - C_{glo}^2\;.$$
Compared with the observed value $C_{ego}$, the constrains from degree assortativities is apparent. Ego networks with negative assortativities all have $C_{ego} < C_{ego}^{rand}$. The exception is Orkut, the only network with positive $r_{ego}$.

\begin{table}
\caption{Assortativities and clustering coefficients at global and ego network levels}
\begin{center}
	\begin{tabular}{|l|l|l|l|l|l|l|l|l|}
	\hline
	\emph{Network} 	& $P_{glo}$		& $C_{glo}$ 	&$C_{ego}^{rand}$	& $C_{ego}$ 	& $r_{glo}$ 		& $r_{ego}$\\
	\hline
	Facebook	& 1.1E-2		& 0.61		&0.848			& 0.76 		& 0.064			& -0.23	\\
	\hline	
	Orkut		& 2.5E-5		& 0.17 		&0.311			& 0.37		& 0.016  		& 0.013\\
	\hline	
	General Relativity& 1.1E-3		& 0.53 		&0.779			& 0.63 		& 0.66 			& -0.14\\
	\hline
	High Energy Physics& 1.6E-3		& 0.61		&0.848			& 0.85		& 0.63			& -0.005\\
	\hline
	Enron email	& 2.7E-4		& 0.50		&0.750			& 0.63		& -0.11			& -0.19\\
	\hline
	LiveJournal	& 4.3E-6		& 0.28		&0.482			& 0.42	 	& 0.045			& -0.248\\
	\hline
	\end{tabular}
\end{center}
\label{tab:clustering_comparison}
\end{table}
In real social networks, the bias of ego networks towards disassortativity and random triangles lead to a ``flattened" view of the global world. If we all build social connections only with local information, assortative cliques are naturally formed even if we try to be open minded. Assortative communities are particularly prevalent in social networks, but this polarization effect is much harder to experience from individual perspectives. Similar lensing effect is also observed in Table~\hyperref[tab:degree_comparison]{\ref*{tab:degree_comparison}}, where the average degrees of neighbors in ego networks $\avg{d_u^v}_{nn}$ decrease from their global counterparts $\avg{d_u}_{nn}$, making the paradox seemingly weaker. While we cannot make precise corrections for degree assortativity and clustering coefficient based on local, incomplete samples, we should always remind ourselves that the former is usually underestimated and the latter actually captures correlations at a higher level than just triangles.

\section{Conclusion}
When only local information is available, statistical perceptions of networks structures deviates systematically from the global ground truth. In this work, we investigate the mathematical relationships between structural features at the global and ego network levels. We proposed a simple approximation of degree distributions of ego networks when the global distribution is known. Combined with empirical observations, we discovered that the heavier tailed degree distribution leads to more disassortative structures and random triangle completion at the ego network level. These insights could help us to better understand and correct for the biases arising from local and limited information in social networks, facilitating more accurate analysis of social behaviors.

\bibliographystyle{splncs03}
\bibliography{Reference}
\end{document}